\documentstyle[aps,multicol,epsf]{revtex}
\input epsf
\draft
\title{Magnetic Properties of a Bose-Einstein Condensate}
\author{M.V. Simkin
and E.G.D. Cohen\\ {\em The Rockefeller University, 
New York, NY 10021-6399} }
\date{ }
\begin{document}
\maketitle
\begin{abstract}
Three hyperfine states of Bose-condensed sodium atoms, recently optically 
trapped,
can be described as a spin-1 Bose gas. We study the behaviour of this 
system in 
a magnetic field, and construct the phase diagram, where the temperature
of the Bose condensation $T_{BEC}$ increases with magnetic field. In particular
the system is ferromagnetic below  $T_{BEC}$ and the magnetization is
proportional to the condensate fraction in a vanishing magnetic field.
Second derivatives of the magnetisation with regard to  temperature 
or magnetic field are discontinuous along the phase boundary.
\end{abstract}
\begin{multicols}{2}
\narrowtext
Recently experimentalists confined three $f=1$ hyperfine states of sodium
in an optical trap \cite{opt}. This was impossible to do in a magnetic trap,
which confines only one hyperfine state \cite{old}. One can ask what are
the properties of  an $f=1$ Bose-Einstein condensate in a magnetic field.
In a weak magnetic field  {\bf H} the electron {\bf s} and nuclear {\bf i} 
magnetic moments 
precess around {\bf f} and {\bf f} precesses around {\bf H}. A simple 
calculation \cite{blokh} gives that the average of the projection of the 
electron spin on the direction of the magnetic field (z-direction) is:
\begin{equation}
\bar{s}_z  = \frac{m_f}{2f(f+1)}[f(f+1)-i(i+1)+j(j+1)].
\label{eq:sz}
\end{equation}
Sodium has a nuclear spin $i=\frac{3}{2}$ and electron spin $s=\frac{1}{2}$, 
and the electron is in
an  $l=0$ state, so that $j=s$. 
Substituting these values in Eq.~(\ref{eq:sz}), we obtain:
\begin{equation}
\bar{s}_z = - \frac{1}{4}m_f.
\label{eq:szmf}
\end{equation}
The magnetic moment of the nucleus is negligible in comparison to that of the
electron. 
Therefore, the system is equivalent to a spin-1 boson, with magnetic moment 4
times less than that of the electron. The projection of the magnetic moment
on z-direction is:
\begin{equation}
\bar{\mu}_z = -2 \mu_B s_z = \frac{\mu_B}{2}m_f,
\label{eq:muzmf}
\end{equation}
where $\mu_B $ is the Bohr magneton. 
As the hyperfine interactions are of the
order of GHz, magnetic field much less then GHz$/\mu_B \approx$100Gs
is small.

As the densities of the Bose condensates of atomic vapours are very small,
one can treat them as an ideal gas. We shall return to the validity of this
assumption below.

An immediate observation is that the system is ferromagnetic\cite{sig}, when 
Bose-condensed, as the condensate is aligned with the external magnetic
field however small it is. Another observation is that the external
magnetic field increases the temperature of the Bose condensation. A spin-$s$
Bose gas has a critical temperature which is $(2s+1)^{2/3}$ times less than 
a spin-0
one at the same density, due to the $2s+1$ spin states available for 
excitations.
At high magnetic fields the gas is completely spin polarized and 
therefore equivalent to a spin-0 Bose gas, and has the same critical 
temperature as the
latter. The aim of this paper is to construct the phase diagram of the system,
to  find the condensate density and magnetization as a 
function of temperature and magnetic field and investigate singularities of the
latter along the phase boundary.

The number of particles in excited states is given by a 
standard formula \cite{llsm},\cite{lon} modified to include 2 extra spin 
states:
\begin{eqnarray}
n_{ex}=\frac{(m_{at}T)^{\frac{3}{2}}}{\sqrt{2} \, \pi^2 \hbar^3} \left (\int_0^{\infty}
\frac{\sqrt{z}dz}{e^{z+a}-1} + \int_0^{\infty}\frac{\sqrt{z}dz}{e^{z+a+y}-1} +
\right. \nonumber \\
\left.\int_0^{\infty}\frac{\sqrt{z}dz}{e^{z+a+2y}-1}\right).
\label{eq:nex}
\end{eqnarray}
Here 
\begin{equation}
y=\frac{(\mu_B/2) H}{T}, a=-\mu/T,
\label{eq:y}
\end{equation}
$\mu$ is the chemical potential and $m_{at}$ is the mass of the bosonic atom.
The integrals in Eq.~(\ref{eq:nex}) can be rewritten following \cite{lon},
\cite{hua}  as 
\begin{eqnarray}
\int_0^{\infty}\frac{\sqrt{z}dz}{e^{z+x}-1}= \frac{\sqrt{\pi}}{2}
\sum_{n=1}^{\infty}\frac{e^{-nx}}{n^{\frac{3}{2}}}= \frac{\sqrt{\pi}}{2}F_{\frac{3}{2}}(x),
\label{eq:ints}
\end{eqnarray}
with $F$  defined as:
\begin{equation}
F_{\alpha}(x)=\sum_{n=1}^{\infty}e^{-nx}/n^{\alpha},
\end{equation}
which reduces to the Riemann function for $x=0$:~ 
$F_{\alpha}(0)=\zeta(\alpha)$.
Let us introduce  new variables:
\begin{eqnarray}
T_0=T_{BEC}(H=0)= \frac{2\pi}{(3 \zeta(\frac{3}{2}))^{2/3}}
\frac{\hbar^2}{m_{at}}n_0^{2/3}, \nonumber \\
t=T/T_0,~~
b=\frac{(\mu_B/2) H}{T_0},
\label{eq:var}
\end{eqnarray}
where $n_0$ is the total particle density.
\begin{figure}
\centering
\begin{minipage}{8.0cm}
\epsfxsize= 8 cm \epsfbox{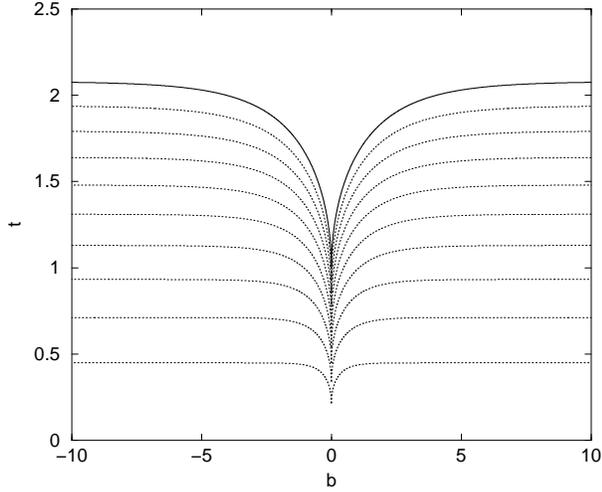}
\end{minipage}
\caption{Contour plot of the Bose-condensate density in $f=1$ sodium atoms 
(spin-1 bosons) in a temperature ($t=T/T_0$) - magnetic field 
($b=\frac{\mu_B H}{2 T_0}$)plane. The solid (and uppermost) line is the
boundary of the phase diagram. The dotted lines are contours for
$n=n_{BEC}/n_0=0.1,0.2,\ldots,0.9$ from top to bottom. Contour lines pile
up near $b=0$ due to the square root singularity in the condensate density 
(see the text). }
\label{fig:phdi}
\end{figure}
In the notation of  
Eq.~(\ref{eq:ints})
and  Eq.~(\ref{eq:var}), Eq.~(\ref{eq:nex}) can be rewritten as
\begin{equation}
n_{ex}/n_0=\frac{t^{\frac{3}{2}}}{3\zeta(\frac{3}{2})} \left (F_{\frac{3}{2}}(a)+F_{\frac{3}{2}}(a+y)+
F_{\frac{3}{2}}(a+2y)\right ).
\label{eq:neqmod}
\end{equation}
Here, as in the familiar spin-0 boson case \cite{llsm}, \cite{lon},
$a=0$ when $n_{ex}(a=0,t,y)/n_0 \le 1$ and $a$ is determined from  
$n_{ex}(a,t,y)/n_0 = 1$ othervise.

The temperature, $t_c$ of BEC is obtained by equating the number of 
particles in 
excited states $n_{ex}$ to the total number of particles: $n_{ex}=n_0$ at
zero chemical potential. Substituting this in  Eq.~(\ref{eq:neqmod})
we obtain:
\begin{equation}
1=\frac{t_c^{\frac{3}{2}}}{3\zeta(\frac{3}{2})} \left (\zeta(\frac{3}{2})+F_{\frac{3}{2}}(y)+
F_{\frac{3}{2}}(2y)\right ).
\label{eq:tofy}
\end{equation}
This gives the phase diagram in a parametric form. In order to obtain it 
explicitly, one  computes numerically $t_c(y)$ from Eq.~(\ref{eq:tofy}) and 
then 
finds $b_c(y)=y/t_c(y)$. To obtain the  condensate density, $n=n_{BEC}/n_0$,  
contour lines, we replace 1 in the left hand side of Eq. \ref{eq:tofy} with 
$1-n$. Then for given $y$ we get $t_n=t_c(1-n)^{2/3}$ and $b_n=y/t_n(y)=
b_c(y)/(1-n)^{2/3}$. Therefore the condensate density contour lines are 
obtained
from the phase transition line by a rescaling.
The results are shown in 
Fig.~\ref{fig:phdi}. For small or large $b$ one can find the
asymptotic behavior of  Eq.~(\ref{eq:tofy}):
\begin{eqnarray}
t_c\sim 1+ \frac{4(1+\sqrt{2})\sqrt{\pi}}{9\zeta(\frac{3}{2})}\sqrt{b}, 
{\rm when}~ b \ll 1, \nonumber \\
t_c \sim 3^{2/3}-\frac{2}{3^{1/3}\zeta(\frac{3}{2})}\exp(-b/3^{2/3}),  
{\rm when}~ b  \gg 1.
\end{eqnarray}
When $b \rightarrow \infty$ all atoms are in the $m_f=1$ state which is
equivalent to having a spin-0 Bose gas at the same density, resulting
in an increase of $t_c$ by a factor of $3^{2/3}$.

\begin{figure}[tb!]
\centering
\begin{minipage}{8.0cm}
\epsfxsize= 8 cm \epsfbox{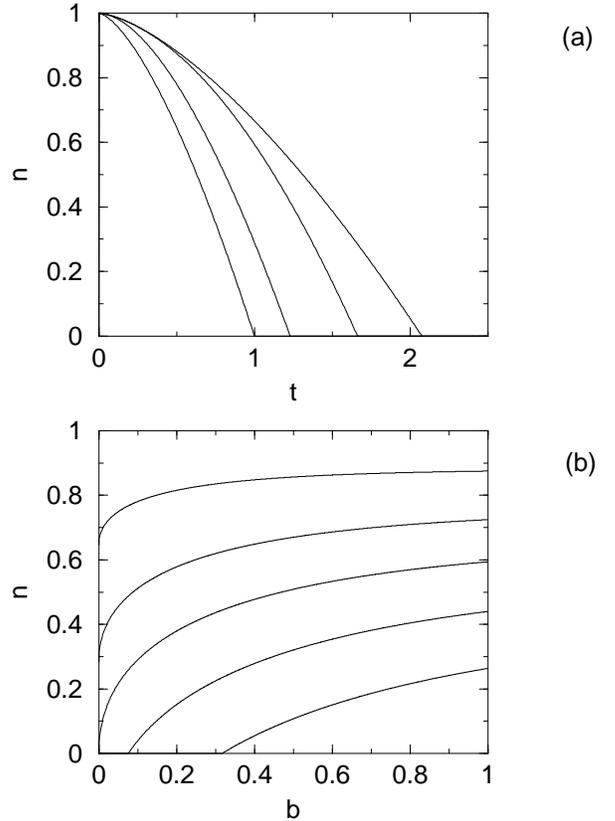}
\end{minipage}
\caption{Fraction of condensate, $n=n_{BEC}/n_0$, as a function of temperature
($t \equiv T/T_0$) and  magnetic field ($b \equiv \frac{(\mu_B/2) H}{T_0}$).
(a) $n(t)$ for several values of $b=0, 0.1, 1, 10$, from left to right.
(b) $n(b)$ for several values of $t=0.5, 0.8, 1, 1.2. 1.4$, from left to 
right.}
\label{fig:odp}
\end{figure}

The condensate fraction $n=n_{BEC}/n_0=1-n_{ex}/n_0$, computed 
using Eq.~(\ref{eq:neqmod}) as a function of magnetic field or temperature
is shown in Fig.~\ref{fig:odp}. 
Asymptotically, the condensate density is $n(b=0)=1-t^{3/2}$ and
$n(b)-n(b=0) \sim t\sqrt{b}$ for $t \le 1$ and $b \ll t$. For $t > 1$,
$n$ is linear in $b-b_c$. 

To calculate the magnetization per particle, $m$, we note that (in units of 
$\frac{\mu_B}{2}$):
\begin{equation}
m=(n(m_f=1)-n(m_f=-1))/n_0.
\label{eq:m0}
\end{equation} 
This expression is 
inconvenient because there are both condensate and excited atoms in the 
$m_f=1$ 
state. Taking into account that $n(m_f=1)+n(m_f=0)+n(m_f=-1)=n_0$, one has 
$m=1-n(m_f=0)-2n(m_f=-1)$, or:
\begin{eqnarray}
m=1- \frac{t^{\frac{3}{2}}}{3\zeta(\frac{3}{2})}
\left (F_{\frac{3}{2}}(a+y)+2F_{\frac{3}{2}}(a+2y)\right ).
\label{eq:m}
\end{eqnarray}
\begin{figure}[tb!]
\centering
\begin{minipage}{8.5 cm}
\epsfxsize= 8.5 cm \epsfbox{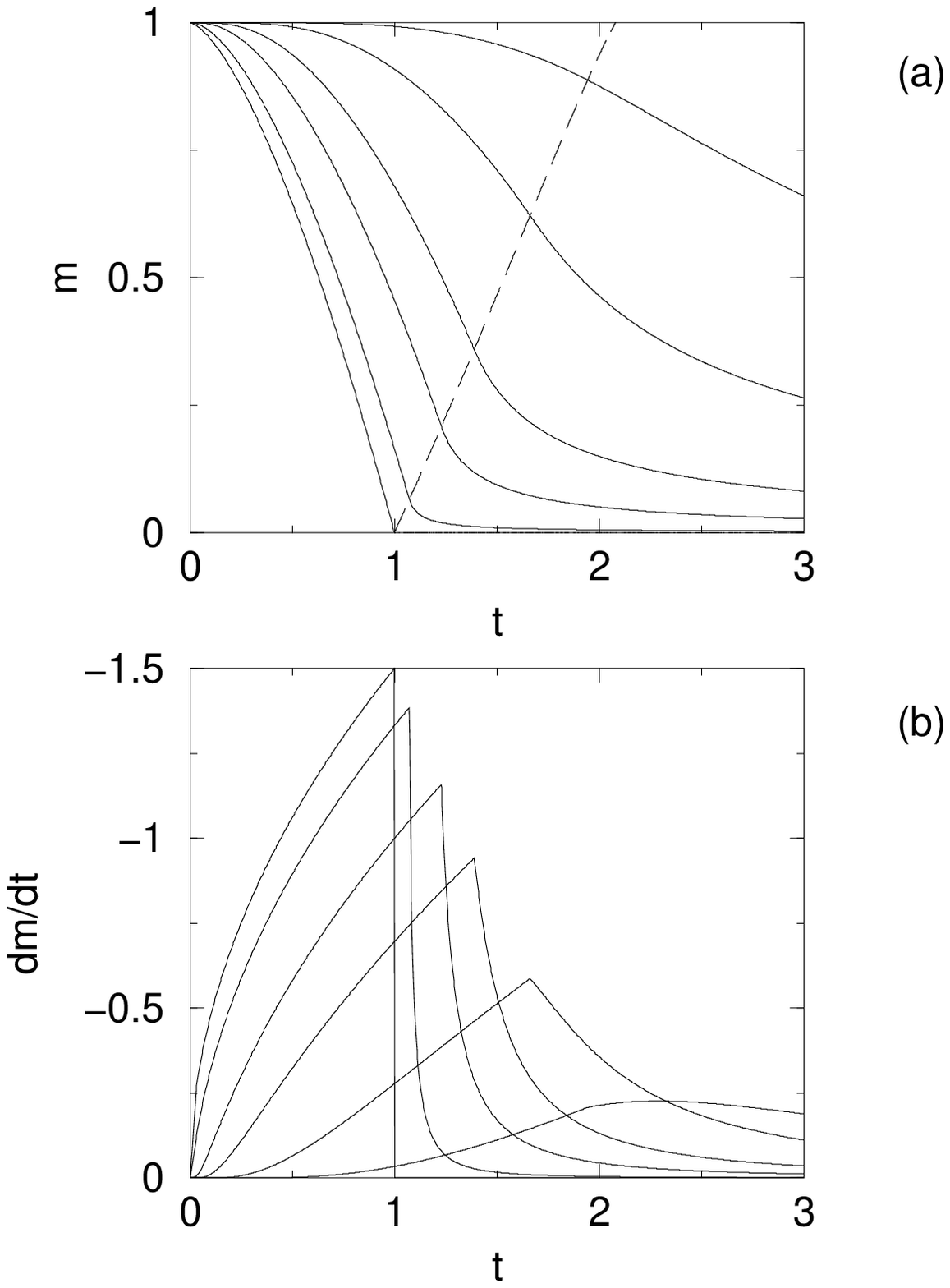}
\end{minipage}
\caption{(a)Magnetization, $m$ , as a function of temperature
($t \equiv T/T_0$) for several values of the magnetic field 
($b \equiv \frac{(\mu_B/2) H}{T_0}= 0, 0.01, 0.1, 0.3, 1, 3$, from left to 
right). Dashed line is $m(b_c(t),t)$.
(b) derivatives of the $m(t)$ curves in (a) with regard to temperature. 
The cusp is at the critical temperature.}
\label{fig:mt}
\end{figure}
Asymptotically:
\begin{equation}
m=1-t^{\frac{3}{2}},
\label{eq:mb0}
\end{equation}
when $b=0+$, that is the magnetisation is proportional to the condensate 
density in a vanishing field, and
\begin{equation}
m \sim \sqrt{b},
\label{eq:mb}
\end{equation}
when $t=1$, {\it i.e.} at the phase transition.

\begin{figure}[tb!]
\centering
\begin{minipage}{8.0cm}
\epsfxsize= 8 cm \epsfbox{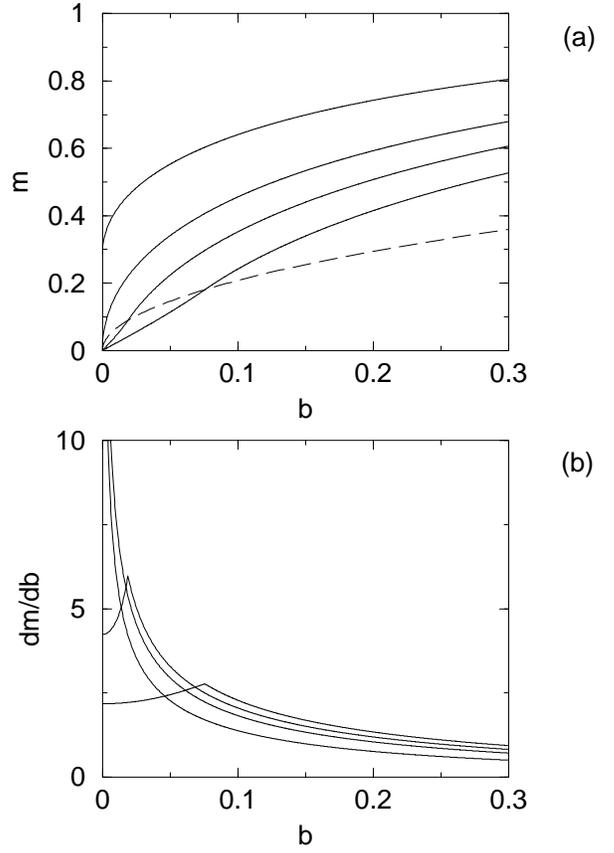}
\end{minipage}
\caption{(a)Magnetization, $m$ , as a function of magnetic field 
($b \equiv \frac{(\mu_B/2) H}{T_0}$)
for severall values of temperature ($t \equiv T/T_0=0.8, 1, 1.1, 1.2$)
from left to right). Dashed line is $m(b,t_c(b))$.
(b) derivatives of the $m(b)$ curves in (a) with regard to magnetic field.
The cusp is at the critical magnetic field. }
\label{fig:mb}
\end{figure}

When $b$ is finite, the magnetization is no longer zero in the normal phase,
however BEC appears in a cusp in the derivatives 
$\left(\frac{\partial m}{\partial t}\right)_b$ and 
$\left(\frac{\partial m}{\partial b}\right)_t$, see Figures \ref{fig:mt}
and \ref{fig:mb}, where these curves were computed numerically,
using equations 
(\ref{eq:dmdt}) and (\ref{eq:dmdb}). In other words,  second derivatives of 
the magnetization
are discontinuous along the phase boundary (see Equations (\ref{eq:d2mdt2})
and (\ref{eq:d2mdb2})).

We have studied the magnetization of the Bose-ferromagnet in a vanishing but 
non zero magnetic field. A question remains what happens for strictly 
zero field. 
In the case of Bose condensate of $N$ spin-1 particles projection of the 
magnetization per particle, $m$ (in units of particle's magnetic moment) 
on the z-axis 
is equal to

\begin{equation}
m_z=(N(1)-N(-1))/N.
\end{equation}
All values of $N(1),N(0),N(-1)$, which satisfy the  obvious restriction
\begin{equation}
N(1)+N(0)+N(-1)=N,
\end{equation}
have equal probability. $m_z=1$ is achieved only when 
$N(1)=N, N(0)=0, N(-1)=0$, while $m_z=0$ can be achieved by many combinations: 
$N(1)=0, N(0)=N, N(-1)=0$, or $N(1)=1, N(0)=N-2, N(-1)=1$ and so on. One can 
show by a generalization of this argument that the probability distribution is
\begin{equation}
P_z(m_z)=1-|m_z|.
\label{eq:pzmz}
\end{equation}

\begin{figure}
\centering
\begin{minipage}{8.0cm}
\epsfxsize= 7 cm \epsfbox{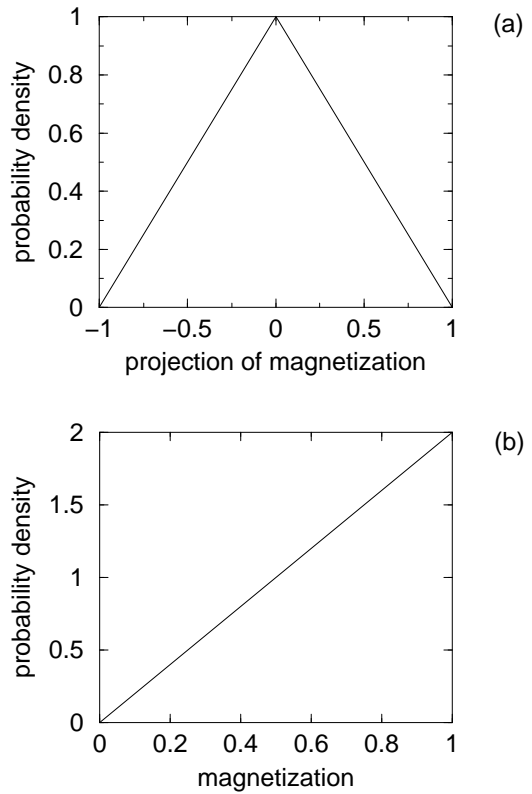}
\end{minipage}
\caption{(a) Probability distribution of the projection on arbitrary axis of 
the spin-1 Bose condensate's magnetization in a zero magnetic field. 
Magnetization is in units of the magnetization of completely spin-polarized 
condensate. (b) Same as (a), but for the absolute value of magnetization 
vector. }
\label{fig:mdist}
\end{figure}

To find how the absolute value of $m$ is 
distributed we notice that the magnetization is macroscopic and therefore can 
be treated classically. Now if we know the distribution of its projection on 
the z-axis $P_z(m_z)$,
we can find the distribution of the absolute value $P(m)$.
Taking into account that the projection of a vector of unit length uniformly 
distributed on a sphere is uniformly distributed between $-1$ and $+1$,
we can write:
\begin{eqnarray}
P_z(m_z)=\int_o^1dm P(m) \frac{1}{2}\int_{-1}^1dx \delta(mx-m_z)= \nonumber \\
\int_{m_z}^1dm P(m)/(2m).
\end{eqnarray}
By differentiating this equation with respect to $m_z$ we get:
\begin{equation}
P(m)=-2m P'_z(m),
\end{equation}
or, taking into account Eq.(\ref{eq:pzmz})
\begin{equation}
P(m)=2m.
\end{equation}
These distributions are shown in Fig. \ref{fig:mdist}.
Let us compare this with classical systems. 
For instanse in the familiar case  of an 
Ising ferromagnet below the phase transition temperature the
probability distribution of the magnetization consists of two one-half delta
functions at positive and negative values of the magnetization. Applying a 
magnetic field changes this distribution to a single delta function. For the 
case of the Heisenberg ferromagnet magnetization can point in any direction, 
however the probability distribution of the absolute value of the 
magnetization 
is a delta function, an infinitely small externall magnetic field will select
the direction of the magnetization. In the case of a Bose-ferromagnet, 
an infinitely small applied magnetic field will not only select the direction,
but also change the form of the probability distribution function
of the magnetization  from that of Fig. \ref{fig:mdist}(b) to delta-function.

The  experiments are done in a trap, which can be modeled by a weak
external potential. It was shown \cite{bay} that interatomic 
interactions, however small,
have a strong effect on the properties of the condensate, when
it is in an external potential. The effect is that the Bose
condensate is not in the  single-particle ground state of the potential,
which has small spatial extent,
but its wave function is spread to balance the interatomic repulsions 
and confining potential. The density of the condensate is then determined from
this balance. Once it is determined, it is a reasonable 
approximation to treat 
the gas as ideal at {\it that} density, as if it was in a square box,
which is in the optical trap
$\sim 10^{15}/{\rm cm}^3$. This  makes $t=1$ in our figures to be of the 
order
of $\mu$K and $b=1$ of the order of $0.1$Gs (which is also much less 
than the weak field requirement of 100Gs mentioned in the begining of the 
paper).
At this density spin-spin interactions are a few orders of magnitude less than
the spin energy in a magnetic field, and can therefore be neglected.
A disadvantage of this low density is that the magnetization is very small 
and difficult to  observe directly. However one can meassure the 
populations of the hyperfine states \cite{opt} and calculate the 
magnetisation from Eq.~(\ref{eq:m0}).
Exchange interactions have been found to be important \cite{ren} in a related 
problem
of a charged spin-0 Bose gas \cite{scha}  (which is diamagnetic when 
Bose-condensed). In our case they also have a profound effect and change the 
order of the transition from second to first.
The previous discussion was of ideal Bose gas in an external magnetic field.
As we argued,  fields of the order of 0.1 Gs (which are also of the order of
the magnetic field of the Earth)  will dominate over exchange
interactions, so the effect of the latter may not be observed. However, if the 
external field is screened down to $10^{-4}$Gs they may play a role.
We shall now address the case of a Bose gas with ferromagnetic exchange 
interactions. Without ferromagnetic exchange the system undergoes a 
conventional Bose 
condensation, while without Bose statistics it undergoes a conventional 
ferromagnetic 
transition. This system is characterized by two parameters:
the Bose condensation temperature $T_{be}$, when ferromagnetic exchange is 
neglected, and the ferromagnetic transition temperature $T_f$, when the 
effects of the
Bose statistics are neglected. The behaviour of the system will depend on the 
ratio $\kappa=T_f/T_{be}$. The details and the complete phase diagram will be 
reported in a future publications, here we shall give an essential result.
If  $\kappa > \kappa_c$, where $\kappa_c$ is close to 1, that is in the case 
of strong ferromagnetic interactions, upon cooling, the system undergoes 
first a
ferromagnetic transition and then at a lower temperature a Bose condensation.
Both transitions are of second order. If  $\kappa < \kappa_c$ then
two transitions merge into a single Bose-ferromagnetic phase transition, which
is of first order. Estimates show that for the condensates, 
experimentalists are currently working with, $\kappa << \kappa_c$. 


A few  final remarks. Since in a vanishing field the condensate carries 
magnetization while the normal
component does not, superflow is accompanied by a transfer of the 
magnetization. In particular, second sound should be accompanied by 
magnetization waves. 
The Einstein-deHaas effect should also be considered. When before
condensation atoms are in equal proportions in $m_f=-1,0,1$ states,
condensation will turn them all into an $m_f=1$ state, which should
result in the  creation of a single vortex.

A spin-1 Bose gas is ferromagnetic when Bose condensed.
The reason for this is the Bose-Einstein statistics, and not the exchange
interactions, like in conventional ferromagnets. Similarly, a charged Bose gas
would behave  like a superconductor\cite{scha}.

The authors acknowlege helpful discussions  with J.T. Liu, H.C. Ren and 
J. Stenger. This work was supported by the Department of Energy  Contract 
No. DE-F602-88-ER13847.


\appendix
\section*{}
From Eq.\ref{eq:m} one obtains:
\begin{eqnarray}
\left(\frac{\partial m}{\partial t}\right)_b =
-\frac{t^{\frac{1}{2}}}{\zeta(\frac{3}{2})} \left[
\frac{1}{2}F_{\frac{3}{2}}(a+y)+F_{\frac{3}{2}}(a+2y)+ \right. \nonumber\\
\frac{y}{3}(F_{\frac{1}{2}}(a+y)+4F_{\frac{1}{2}}(a+2y)) - \nonumber\\ \left.
\frac{1}{3}(F_{\frac{1}{2}}(a+y)+2F_{\frac{1}{2}}(a+2y))t
\left(\frac{\partial a}{\partial t}\right)_b \right] .
\label{eq:dmdt}
\end{eqnarray}
Where $\left(\frac{\partial a}{\partial t}\right)_b$ can be found from 
Eq.(\ref{eq:neqmod}) to be:
\begin{eqnarray}
t\left(\frac{\partial a}{\partial t}\right)_b=\left[\frac{3}{2}
(F_{\frac{3}{2}}(a)+F_{\frac{3}{2}}(a+y)+F_{\frac{3}{2}}(a+2y))+ 
\right. \nonumber\\ \left.
y(F_{\frac{1}{2}}(a+y)+2F_{\frac{1}{2}}(a+2y))\right]\nonumber\\
\left[F_{\frac{1}{2}}(a)+F_{\frac{1}{2}}(a+y)+F_{\frac{1}{2}}(a+2y)
\right]^{-1}.
\label{eq:dadt}
\end{eqnarray}
Now $\left(\frac{\partial m}{\partial t}\right)_b$ can be found numerically 
from Eqs.
(\ref{eq:dmdt}), (\ref{eq:dadt}), with $a$  from Eq.(\ref{eq:neqmod}).
To find the discontinuity in the $\frac{\partial^2 m}{\partial t^2}$
at the phase transition we note that when $a=0$: 
$\left(\frac{\partial a}{\partial t}\right)_b=0$ (Cf.Eq.(\ref{eq:dadt}),
taking into account that $F_{\frac{1}{2}}(0)=\infty$). Therefore the
discontinuity comes only from the second derivative of $a$. We find for the
jump:
\begin{eqnarray}
\Delta\frac{\partial^2 m}{\partial t^2}=
\frac{F_{\frac{1}{2}}(y)+2F_{\frac{1}{2}}(2y)}
{6\pi\zeta(\frac{3}{2})\sqrt{t}} 
\left[\frac{3}{2}(\zeta(\frac{3}{2})+F_{\frac{3}{2}}(y)+
\right.\nonumber\\ \left.
F_{\frac{3}{2}}(2y))+ 
y(F_{\frac{1}{2}}(y)+2F_{\frac{1}{2}}(2y))\right]^{2}.
\label{eq:d2mdt2}
\end{eqnarray}
This has assymptotics $\Delta\frac{\partial^2 m}{\partial t^2} 
\sim b^{-\frac{1}{2}}$ for $b \ll 1$ and  $\Delta\frac{\partial^2 m}
{\partial t^2} \sim \exp(-b/3^\frac{2}{3})$ for $b \gg 1$.

Derivatives with regard to the magnetic field can be done analogously, leading
to:
\begin{eqnarray}
\left(\frac{\partial m}{\partial b}\right)_t =
\frac{t^{\frac{1}{2}}}{3\zeta(\frac{3}{2})} \left[
F_{\frac{1}{2}}(a+y)+4F_{\frac{1}{2}}(a+2y)) + \right. \nonumber\\ \left.
(F_{\frac{1}{2}}(a+y)+2F_{\frac{1}{2}}(a+2y))t
\left(\frac{\partial a}{\partial b}\right)_t \right] .
\label{eq:dmdb}
\end{eqnarray}

\begin{eqnarray}
t\left(\frac{\partial a}{\partial b}\right)_t=-\frac{F_{\frac{1}{2}}(a+y)+
2F_{\frac{1}{2}}(a+2y)}{F_{\frac{1}{2}}(a)+F_{\frac{1}{2}}(a+y)+
F_{\frac{1}{2}}(a+2y)}.
\label{eq:dadb}
\end{eqnarray}

\begin{eqnarray}
\Delta\frac{\partial^2 m}{\partial b^2} = 
-\frac{\left[F_{\frac{1}{2}}(y)+2F_{\frac{1}{2}}(2y)\right]^3}
{6\pi\zeta(\frac{3}{2})\sqrt{t}}. 
\label{eq:d2mdb2}
\end{eqnarray}
This has assymptotics $\Delta\frac{\partial^2 m}{\partial b^2} 
\sim -\frac{1}{(t-1)^3}$ for $t-1 \ll 1$ and  $\Delta\frac{\partial^2 m}
{\partial b^2} \sim -(3^\frac{2}{3}-t)^3$ for $3^\frac{2}{3}-t \ll 1$.

One can show similarly that $\frac{\partial^2 m}{\partial t \partial b}$ is 
also discontinuous along the phase boundary.

\end{multicols}
\end{document}